\documentclass[referee]{raa} 
\UseRawInputEncoding
\usepackage{amssymb,amsmath}
\usepackage{mathrsfs}
\usepackage[papersize={9.0in,14in}]{geometry}
\usepackage{txfonts}
\usepackage{hyperref}
\hypersetup{
    colorlinks=true,
    linkcolor=blue,
    citecolor=blue,
    urlcolor=blue}
\usepackage{graphicx,times}             
\usepackage{natbib}
\usepackage{booktabs}
\bibpunct{(}{)}{;}{a}{}{,}
\begin{document} 

   \title{A Study of Subsurface Convection Zones of Fast Rotating Massive Stars}
   \volnopage{Vol.0 (20xx) No.0, 000--000}      
   \setcounter{page}{1}          
	\author{Xiaolong He \inst{1,2}
\and Guoliang L\"{u} \inst{1,2}\thanks{Corresponding author:guolianglv@xao.ac.cn}
\and Chunhua Zhu\inst{1}\thanks{Corresponding author:chunhuazhu@sina.cn}
\and Lin Li\inst{1}
\and Helei Liu\inst{1}
\and Sufen Guo\inst{1}
\and Xizhen Lu\inst{1}
\and Lei Li\inst{1}
\and Hao Wang\inst{1}}

   \institute{School of Physical Science and Technology, Xinjiang University, Urumqi 830046, People's Republic of China {\it chunhuazhu@sina.cn}\\
              \and 
              Xinjiang Astronomical Observatory, Chinese Academy of Science, 150 Science 1-Street, Urumqi 830011, Peopleʼs Republic of China {\it guolianglv@xao.ac.cn }\\
\vs\no
   {\small Received 20xx month day; accepted 20xx month day}
              }


\abstract{The subsurface convective zones (CZs) of massive stars significantly influences many of their key characteristics. Previous studies have paid little attention to the impact of rotation on the subsurface convective zone (CZ), so we aim to investigate the evolution of this zone in rapidly rotating massive stars. We use the Modules for Experiments in Stellar Astrophysics (MESA) to simulate the subsurface CZs of massive stars during the main sequence phase. We establish stellar models with initial masses ranging from 5 \,$M_{\odot}$ to 120 \,$M_{\odot}$, incorporating four metallicities ($Z$ = 0.02, 0.006, 0.002, and 0.0001) and three rotational velocities ($\mit\omega/\omega_{\text {crit}}$ = 0, $\mit\omega/\omega_{\text {crit}}$ = 0.50, and $\mit\omega/\omega_{\text {crit}}$ = 0.75). We find that rapid rotation leads to an expansion of the subsurface CZ, increases convective velocities, and promotes the development of this zone. Additionally, subsurface CZs can also emerge in stars with lower metallicities. Comparing our models with observations of massive stars in the Galaxy, the Large Magellanic Cloud, and the Small Magellanic Cloud, we find that rotating models better encompass the observed samples. Rotation significantly influences the evolution of the subsurface CZ in massive stars. 
By comparing with the observed microturbulence on the surfaces of OB stars, we propose that the subsurface CZs may be one of the sources of microturbulence.  
\keywords{stars: massive  ---  stars: evolution, turbulence, convection  ---  stars: early-type ---  stars: rotation}
}
  \authorrunning{X.-L. He et al.}
  \titlerunning{Subsurface CZs of Fast Rotating Massive Stars}
\maketitle
\section{Introduction}\label{sec:introduction}

Massive stars ($M$ \,$\geq$\, 8\,$M_{\odot}$) take an important role in many fields of modern astrophysics \citep{Kobayashi2006,2012Langer}. For example,
at the time of their death, they eject large amounts of heavy elements into the universe (supernova explosions). This process affects the surrounding interstellar medium (ISM) \citep[e.g.,][]{2013Kippenhahn,2014Hopkins,2016Crowther,2016Stark,2019Yu,2020BowmanISM,2021Yu,2021Wu}, although this is not the only way (e.g., ejecta of classical novae, binary mergers, etc. \citep[e.g.,][]{2013Lv,2013Zhu,2016Li,2017Rukeya,2019Zhu,2019Duolikun,2020Shi,2020Lv}), even if other contributions are smaller. Massive stars, also known as cosmic engines, have relatively short lifetimes ($10^6$ - $10^8$ years) because massive stars are rich sites for nucleosynthesis.

Massive stars are crucial objects. Through advancements in observational tools and theoretical simulations, fundamental concepts about their evolution and structure have been established. However, current stellar models have yet to fully comprehend certain observed properties of massive stars (e.g., surface magnetism and bright spots \citep{2014Ramiaramanantsoa,2019Cantiello,2023Henriksen}; mass-discrepancy \citep{2015Markova,2021Serenelli}; envelope inflation \citep{2023Lu}; surface turbulence \citep{2009Cantiello,2014Simon-Diaz,2017Godart}; stochastic low-frequency photometric (SLF) variability \citep{2011Blomme,2019Pedersen,2019NatAs...3..760B,2019Bowman,2020Bowman,2021Rauw,2021Cantiello,2023Shen} ). These highlight the necessity for a more comprehensive understanding of various physical processes in massive stars.

Reviewing the history of theoretical simulations of massive stellar structures, \citet{1993Stothers} predicted the existence of a small convective zone (CZ) in the envelope of sufficiently luminous massive main sequence models. This completes the Cowling model proposed previously, and \citet{2009Cantiello} refers to these small CZs as subsurface CZs. The subsurface CZ is divided into HI/HeI/HeII/Fe CZs depending on recombinant particles \citep{2022JermynAtlas}.

Recent studies of the subsurface CZ in massive stars have garnered considerable interest \citep{2009CoAst.158...61C,2009Cantiello,2011Cantiello3DMHD,2011IAUS..273..200C,2021Cantiello,2011Cantiello,2019Cantiello,
2015Grassitelli,2015ApJ...808L..31G,2016A&A...590A..12G,2016Grassitelli,2018Grassitelli,2021Grassitelli,2022JermynWindow,
2022JermynAtlas,2020Jermyn,2021JermynMagnetic,2019Lecoanet,2015Jiang,2017Jiang,2018Jiang}. These investigations suggest that the characteristics of the CZ may elucidate various phenomena observed on stellar surfaces \citep[e.g.,][]{2009Cantiello,2010Fraser,2015Grassitelli,2016Grassitelli,2017Simon-Diaz,2021Cantiello,2022Schultz}. \citet{2009Cantiello} studied these subsurface CZs potential connections with microturbulence in massive stars, and showed the corresponding observational evidence. \citet{2015Grassitelli,2016Grassitelli} argued that in massive stars, macroturbulent velocities and the pressure induced by the turbulent motion in subsurface CZs are related and discussed their dependence on metallicity. \citet{2022JermynAtlas} classified the ionization-driven subsurface CZs into H I/He I/He II/Fe CZs according to temperature and gave an atlas of convection in main sequence (MS) stars. And the subsurface CZs could also turn out to directly affect the evolution of stars, by influencing the mass loss rate \citep{2011Cantiello3DMHD}.

However, the previous studies rarely consider the rotation's effects on the subsurface CZs. It is well known that the rotation is actually essential for the formation and evolution of MS stars \citep{2000Maeder,2003Meynet,2004Aerts,2005Maeder,2010Maeder,2012Langer,2014Maeder,2023Zhu,2023Abdul-Masih}. Rotation not only modifies the evolutionary trajectory, lifetime, and final product of these stars, but is also thought to induce a chemically homogeneous evolution (CHE) that allows the transport of core-processed elements to the surface \citep{1987Maeder,2016Mandel,2023Wang,2023Lilei,2024Chen,2020Han}. Some of the larger spectroscopic observational surveys: the two surveys using the Fibre Large Array MultiElement Spectrograph (FLAMES) on the Very Large Telescope (VLT) \citep{2005Evans,2011Evans}, the IACOB project \citep{2011Simon-Diaz,2014Simon-Diaz,2015Simon-Diaz} and the Large Sky Area Multi-Object Fiber Spectroscopic Telescope (LAMOST, also known Guo Shou Jing telescope) survey \citep{2012Zhao,2012Deng,2012Cui,2019Liu,2019Chen,2020Wang,2021YuLAMOST,2021Li,2022Yan,2023Li}, have made significant contributions to our understanding of massive stars. These surveys have provided quantitative analyses \citep{2018Holgado,2022Holgado} of stellar parameters such as rotational velocities ({$v \sin i$\,}) and additional line broadening effects \citep{2013Dufton,2013Ramirez-Agudelo,2015Ramirez-Agudelo,2018Dufton,2014Simon-Diaz,2018Markova}. The {$v \sin i$\,} distribution of stars, as revealed by previous studies, exhibits a bimodal character \citep[e.g.,][]{1977Conti,2013Ramirez-Agudelo,2023Henriksen}, with tails of fast rotators extending up to $450 {\, {\rm km}\, {\rm s}^{-1}}$ \citep{2014Simon-Diaz,2022Holgado}. For massive stars, rotation velocities ({$v \sin i$\,}) exceeding $150 {\, {\rm km}\, {\rm s}^{-1}}$ are considered fast rotations \citep{2013deMink}, while those surpassing the threshold of $300 {\, {\rm km}\, {\rm s}^{-1}} $ are deemed extreme rotations \citep{2022Holgado}.

\citet{2008Maeder} suggested that the latitudinal structure of the subsurface CZs is substantially altered at the highest rotation rates. On the contrary,
some researchers say that the influence of rotation on the subsurface CZ is not enough to be paid attention to \citep{2009Cantiello}. Based on the classification method in \citet{2022JermynAtlas}, we investigate the effects of varying metallicities and rotation rates on the subsurface CZ. By analyzing variations in average convective velocities within the CZ and comparing them to observed microturbulent velocities, we seek to enhance our understanding of the formation mechanisms of microturbulence in massive stars, providing a more comprehensive explanation for stellar evolution and observational phenomena.

This paper is organized as follows. We present our modeling approach in Section~\ref{sec:two}.
In Section~\ref{sec:three}, we compare the rotational and non-rotational models given in the modeling calculations, give the effect of rotation on the subsurface CZs, and compare the models with the observations and discuss them. Finally, conclusions are given in Section~\ref{sec:four}.
\begin{figure}
  \centering
  \resizebox{\hsize}{!}{\includegraphics{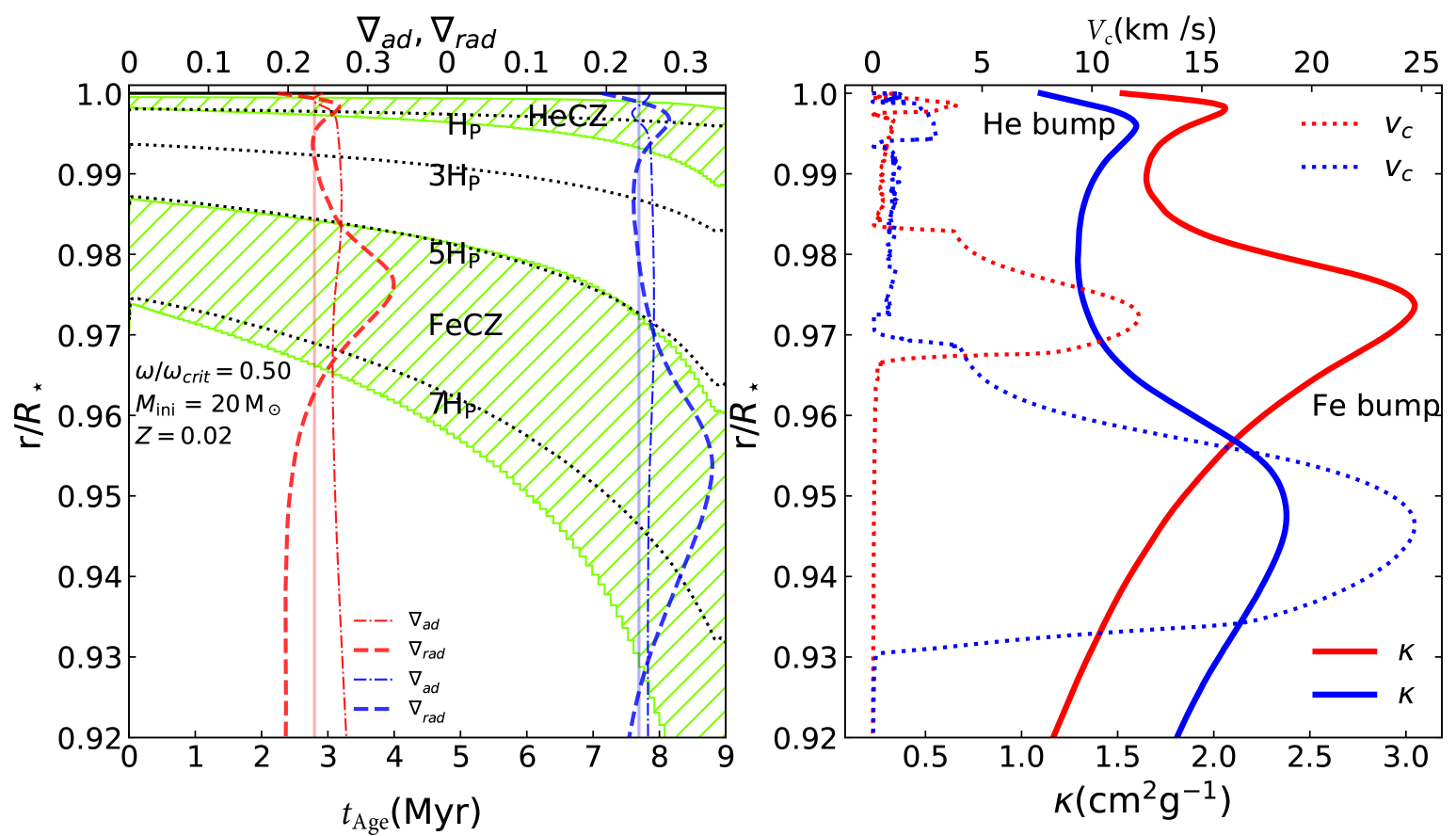}}
  \caption{
  The internal structure of 20\,$M_{\odot}$ star with $Z$ = 0.02 and $\mit\omega/\omega_{\text {crit}}$ = 0.50 around the stellar surface 
  within 8$\%$ of stellar radius ($R_{\star}$). The left panel is Kippenhahn diagram from zero-age main sequence to the end of core hydrogen burning.  
  The subsurface CZs are represented by green line regions, and the dotted lines give the positions of $nH$$_{\rm P}$ \citep{2021Cantiello}. 
  Simultaneously, the adiabatic gradient $\nabla_{\rm ad}$ and the radiative gradient $\nabla_{\rm rad}$ at $t_{\text {Age}}$ $\approx$ 2.8 and 7.7 Myr are given by red and blue dash-dotted and dashed lines, respectively. Their values are labeled by the top axis. 
  The right panel shows the opacity $\kappa$ (X axis) and the convective velocity ${V}_{\rm c}$ (top axis) by solid and dashed lines when  $t_{\text {Age}}$ $\approx$ 2.8 (red) and 7.7 Myr (blue), respectively. The opacity bumps caused by iron and helium ionization are marked.
  }
  \label{fig:20M_0.02z_0.5w_image}
\end{figure}

\section{Models}\label{sec:two}
We utilize the Modules for Experiments in Stellar Astrophysics (MESA; version15140, \citep{2011Paxton, 2013Paxton,2015Paxton,2018Paxton,2019Paxton}) to construct a grid of stellar evolution models with initial masses ranging from 5 \,$M_{\odot}$ to 120 \,$M_{\odot}$ \footnote{We compute stellar evolution models for the following initial masses:
5.0, 5.2, 5.4, 5.6, 5.8, 6.0, 6.5, 7.0, 7.5, 8.0, 9.0, 10, 11, 12, 13, 14, 15, 16, 17,
18, 19, 20, 21, 22, 23, 24, 25, 30, 40, 50, 60, 80 and 120\,$M_{\odot}$.} 
and metallicities of $Z$ = 0.02, $Z$ = 0.006, $Z$ = 0.002 and $Z$ = 0.0001, representing approximately the Galaxy, Large Magellanic Cloud (LMC), Small Magellanic Cloud (SMC), and extremely metal-poor (EMP) stars, respectively. Additionally, we focus on rotation rates of $\mit\omega/\omega_{\text {crit }}$ = 0.50 and $\mit\omega/\omega_{\text {crit }}$ = 0.75 ($\mit\omega$ and $\mit\omega_{\text {crit }}$ represent the angular velocity and critical angular velocity, respectively.). The convection boundary is calculated using the Ledoux criterion,
and the mixing-length parameter $\alpha_{\mathrm{LMT}}$ = 1.5 is adopted \citep{2009Cantiello,2011Brott,2017Lv,2017Zhu,2018Wang,2018Cui,2022Long}.
Following \citet{2022JermynAtlas}, we employ convective premixing ~\citep[][Section 5.2]{2019Paxton} and Cox MLT option \citep{1968Cox}.
We use the same step-function overshooting model as \citet{2021Cantiello}
to calculate the overshoot area,
where the core overshooting is applied in the exponential scheme with
parameters $f_{\rm ov} = 0.014$ and $f_{0} = 0.004$ \citep{2020Shi}.
As in the work of \citet{2009Cantiello},
our model uses nuclear reaction rates from JINA REACLIB \citep{2010Cyburt},
with radiative opacities mainly from OPAL \citep{1993Iglesias,1996Iglesias}, and electron conduction opacities from \citet{2007Cassisi}. Similarly, our calculations are based on the provisions of \citet{2001Vink} for mass loss due to stellar wind. We adopt the \texttt{approx21\_cr60\_plus\_co56.net} nuclear network in MESA, which includes 22 nuclides.

Following the study of \citet{2009Cantiello},
we define the average of a general quantity $q$ as
\begin{equation}
\bar{q} \equiv \frac{1}{\alpha_{\mathrm{MLT}} \mathrm{H}_{\mathrm{P}}} \int_{\mathrm{R}_{\mathrm{c}}-\alpha_{\mathrm{MLT}} \mathrm{H}_{\mathrm{P}}}^{\mathrm{R}_{\mathrm{c}}} q(r) d r,
\label{Eq:FeCZ_aver_V_conv}
\end{equation}
where $H$$_{\rm P}$ is the pressure scale height calculated
at the upper boundary (${\rm R}_{\rm c}$) of the subsurface CZ. 
Using the aforementioned method, we compute the radial average of quantity $q$ over the distance $\alpha_{\mathrm{LMT}}$$H$$_{\rm P}$ downward from the upper boundary Rc of the CZ. We focus on the upper part of the CZ (near the surface) due to our interest in observable surface phenomena.
\citet{2022JermynAtlas} defined the radial average of a quantity $q$ over a CZ by
\begin{equation}
\begin{aligned}
	\langle q \rangle \equiv \frac{1}{\delta r}\int_{\rm CZ} q dr,
\label{Eq:HeCZ_aver_V_conv}
\end{aligned}
\end{equation}
where $\delta r$ is the thickness of the CZ.
These are intended to quantitatively analysis the parameters of the CZ,
but they depend on the quantities that vary within the CZ itself,
so we need average values.

Using the approach of \citet{2022JermynAtlas}, we distinguish the subclasses  of thin ionization-driven convective regions (subsurface CZs) based on temperature. 
In general,
the thin ionization zones develop at almost fixed temperatures,
so the temperature can be a good representative of the ionization state in this region.
Fig. \ref{fig:20M_0.02z_0.5w_image} shows the evolution of the normalized radial positions of the helium convection zone (HeCZ) and the iron convection zone (FeCZ) of the 20 \,$M_{\odot}$ model during the MS ($\sim$ 9 Myr) (compare to Fig.1 of \citet{2021Cantiello},
where the HeCZ and the FeCZ have the same evolutionary trend).
Opacity $\kappa$ is a function of temperature (as shown in Fig.1 of \citet{2009Cantiello}). 
It is very helpful for understanding the evolution of the subsurface CZ over time in terms of its radial position:
as the star evolves and its surface temperature decreases, the subsurface CZ moves inward, as indicated by the red and blue solid lines, with the opacity peak shifting radially inward. This fact would be from the fact that the Fe bump caused by the Fe transition lines appears at a temperature of Log $T$ $\approx5.3$ \citep{2009Cantiello}. The temperature region moves inward with the expansion of the star. 
We then show the profiles of the variation of the opacity and
convective velocity of the model at the early ($\sim$ 30$\%$) and late ($\sim$ 86$\%$) stages of the MS evolution
in normalized radial positions.
We obtain that the model has a helium (He) opacity peak at the beginning of the evolution,
at higher effective temperatures,
in the very near stellar surface region,
whose presence gives rise to the convective velocity bump illustrated
by the dotted line on the right in Fig. \ref{fig:20M_0.02z_0.5w_image}.
Simultaneously, we can see the same radial position,
the consistency of the iron (Fe) opacity peak and the convective velocity.
Compared to the FeCZ, the HeCZ is thinner and weaker, so we primarily focus on the FeCZ. 
Additional, one should note that the presence of the subsurface CZ is also
related to other factors, such as temperature gradients (as shown in the left panel of Fig. \ref{fig:20M_0.02z_0.5w_image},
with the dashed line indicating the radiative gradient $\nabla_{\rm rad}$
and the dash-dotted line indicating the adiabatic gradient $\nabla_{\rm ad}$), which will be further discussed in the next section.

\begin{figure}
  \centering
  \resizebox{\hsize}{!}{\includegraphics{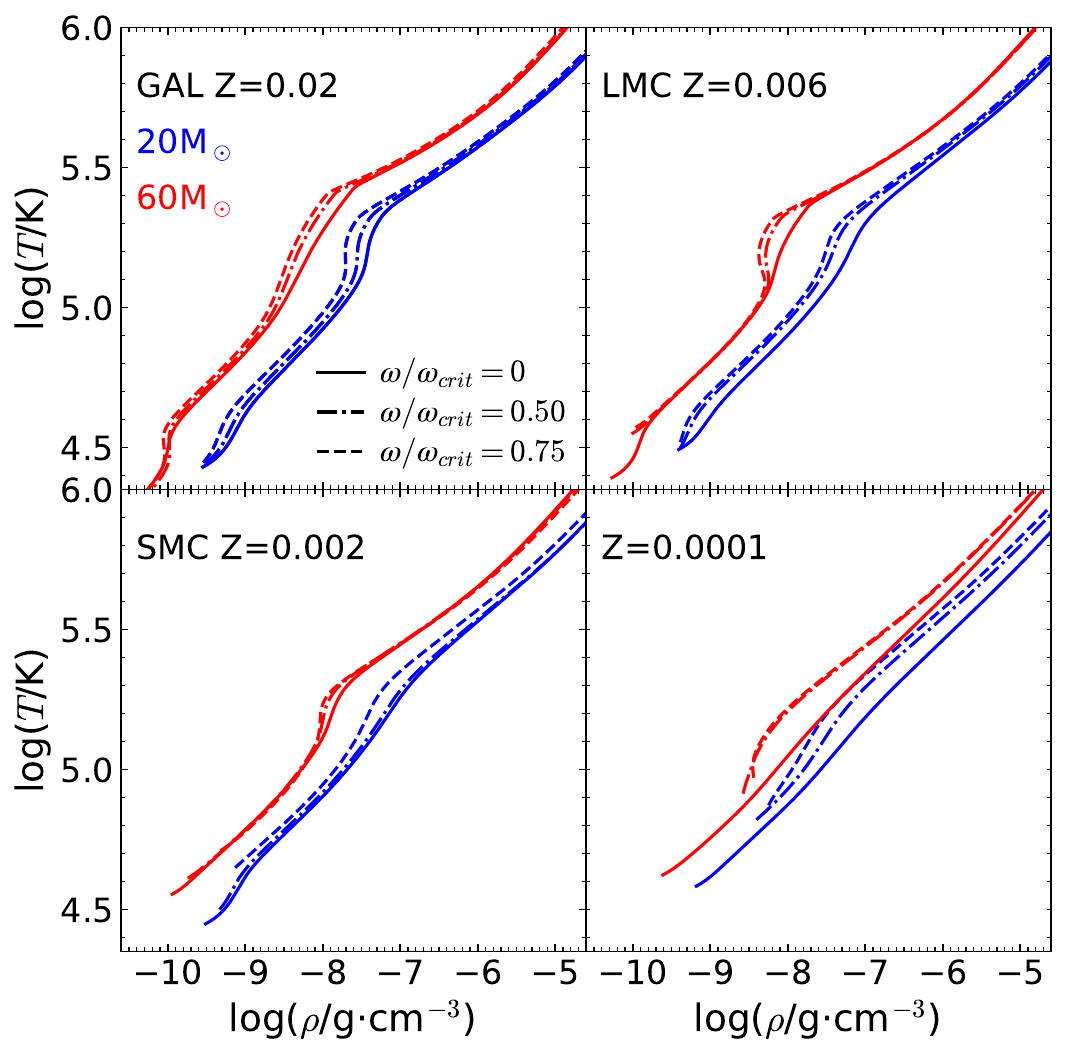}}
  \caption{
  The stellar profile of mass density and temperature at the end of the MS for different metallicities ($Z$ = 0.02, 0.006, 0.002 and 0.0001). 
  The blue and red curves show the profiles of
  the 20 \,$M_{\odot}$ and 60 \,$M_{\odot}$ ZAMS stars.
  The different lines show the three rotational velocities:
  the solid line shows the model without rotation,
  the dash-dotted and dashed lines show the models with $\mit\omega/\omega_{\text {crit }}$ = 0.50
  and $\mit\omega/\omega_{\text {crit }}$ = 0.75, respectively.
  }
  \label{fig:T_Rho_image}
\end{figure}

\section{Results}\label{sec:three}
According \citet{2008Maeder},
their 2D simulations suggest that rotation facilitates the development of subsurface CZs near the equator of massive stars.
In our 1D simulations, Fig. \ref{fig:T_Rho_image} illustrates the temperature-density profiles of rotating and non-rotating models at the end of the MS to highlight the effects of rotation.
As shown in Fig. \ref{fig:T_Rho_image}, a small bulge appears at Log $T$ $\approx5.3$, a fluctuation in density that decreases in this region as the rotation rate increases (see also Table ~\ref{table}).
The density inversion in this small region is the presence of a CZ triggered by the Fe opacity peak.
Another small bulge is at Log $T$ $\approx4.7$, which is triggered by the He opacity peak, closer to the stellar surface.
At the same metallicity, the stellar surface temperature increases with increasing rotational velocity. This is due to rotation enhancing the nuclear burning region, resulting in higher luminosity and surface temperature \citep{2011Brott,2012Ekstr1}. 
Simultaneously, as the metallicity decreases,
the surface temperature of the star increases to the point where the CZ triggered
by the He opacity peak disappears ($Z$ = 0.0001). 
This is due to the reduction in opacity as stellar metallicity decreases. Therefore, comapred to the star with solar metallicity, a low-metallicity star has smaller radius and higher effective temperature, which results in the subsurface temperature exceeding the He ionization temperature \citep{2011Brott,2022JermynAtlas}. 
Comparing the temperature-density profiles of 20 \,$M_{\odot}$ and 60 \,$M_{\odot}$ stars, the increase in initial stellar mass results in a larger radius, leading to a lower density at the surface. Additionally, as the rotational velocity increases, the small peaks become more pronounced, as rotation enhances the radiative gradient, which will be explained in further detail below. 
We should note that the density of the above regions is extremely low ($\rho\approx 10^{-10} \sim 10^{-8} \mathrm{g} \cdot \mathrm{cm}^{-3}$), so the convective transmission is very inefficient and the energy transfer is dominated by radiation.
These mean that rotation dramatically changes the profile of the stellar subsurface and around the surface \citep{2008Maeder}. This is because the radiation gradient in a rotating star can be written as
\begin{equation}
\nabla_{\mathrm{rad}}=\frac{\Gamma}{4(1-\beta)\left(1-\frac{\mit\omega^2}{2 \pi G \rho_{\mathrm{m}}}\right)}
\end{equation}
where $\Gamma$ is the Eddington factor, $\beta= P_{\mathrm{g}}/P$ is the ratio of gas pressure to total pressure, $\mit\omega$ is the angular velocity, $G$ is the gravitational constant, and $\rho_{\rm{m}}$ is the average density inside the considered isobar.  Here the above equation $(1-\frac{\mit\omega^2}
{2 \pi G \rho_{\rm{m}}})$ = $M_{\star}(r)/M_r$ represents the ratio of the effective mass to the mass inside the isobar, i.e., the rate of change of the mass reduced by the centrifugal force. We can see that all terms are localized. As shown in Fig. \ref{fig:T_Rho_image}, the fluctuations in density and temperature in a rotating star make the effective mass $M_{\star} < M$, which results in a larger radiation gradient in a rotating star compared to a non-rotating star. This result is important and supports the idea that rotation favors convection.

In the following subsections, 
we discuss the effects of rotation on these subsurface CZs
and comparison with the observations.
\begin{table*}
\begin{minipage}[t]{\textwidth}
\centering
\caption{
  Properties of FeCZs in our 20 \,$M_{\odot}$ models of $Z$ = 0.02. These models also are shown in the top panel of Fig.\ref{fig:kipp_plot_20M_image} (a, a1, a2). The values in the table, from up to down, corresponding in order to the rotation parameters $\mit\omega/\omega_{\text {crit }}$ = 0, 0.50 and 0.75. The ages are $\sim$85$\%$ of the MS lifetime, $7.00\times10^{6}$ yr, $7.77\times10^{6}$ yr and $8.17\times10^{6}$ yr, respectively.
  }
\label{table}
\renewcommand{\footnoterule}{}
\begin{tabular}{l| c c c c c c c c c c c c c c}
\hline\hline
   $M_{\text {init} }$&
   $\mit\omega/\omega_{\text {crit }}$&
   $R_{\star}$&
   $R_{\mathrm{FeCZ}}$\footnote{Radial coordinate of the top of the FeCZ.}&
   $\Delta R_{\mathrm{FeCZ}}$\footnote{Radial extension of the FeCZ.}&
   ${\mit{H}_{\mathrm{P}}}$\footnote{Pressure scale height at top/bottom of the FeCZ.}&
   ${V}_{\rm c}$\footnote{Maximum of the convective velocity inside the FeCZ.}&
   log $\rho$\footnote{Density at ${V}_{\rm c}$.}&
   log $M_{\rm FeCZ}$\footnote{Mass contained in the convective region.}&
   log $M_{\rm top}$\footnote{Mass in the radiative layer between the stellar surface and the upper boundary of the CZ.}&
   $\tau_{\rm turn}$\footnote{Convective turnover time, $\tau_{\rm turn} = {\mit{H}_{\mathrm{P}}}/{V}_{\rm c}$.}&
   $\tau_{\rm conv}$\footnote{Time that a piece of stellar material spends inside a convective region, $\tau_{\rm conv} = \Delta M_{\rm FeCZ}/\dot M$.}&
   $\dot{M}$ \\
   \,$M_{\odot}$&
   &
   ${\,{\rm R}_\odot}$&
   ${\,{\rm R}_\odot}$&
   ${\,{\rm R}_\odot}$&
   ${\,{\rm R}_\odot}$&
   km~s$^{-1}$&
   g~cm$^{-3}$&
   \,$M_{\odot}$&
   \,$M_{\odot}$&
   days&
   days&
   \,$M_{\odot}$ $\rm{ yr}^{-1}$ \\

   \hline
     & 0   & 11.67& 11.37& 0.34& 0.090-0.270& 15.34& -7.24& -5.310& -6.137& 0.26& 25156& $7.1\times10^{-8}$ \\
   20& 0.5 & 11.86& 11.54& 0.47& 0.098-0.319& 24.69& -7.31& -5.261& -6.102& 0.08& 17412& $1.2\times10^{-7}$ \\
     & 0.75& 11.28& 10.97& 0.54& 0.083-0.322& 26.60& -7.35& -5.273& -6.344& 0.08& 13429& $1.6\times10^{-7}$ \\

   \hline

\end{tabular}
\end{minipage}
\end{table*}

\subsection{Effect of rotation on the subsurface CZs}\label{sec:three_effect}
\begin{figure}
  \centering
  \resizebox{\hsize}{!}{\includegraphics{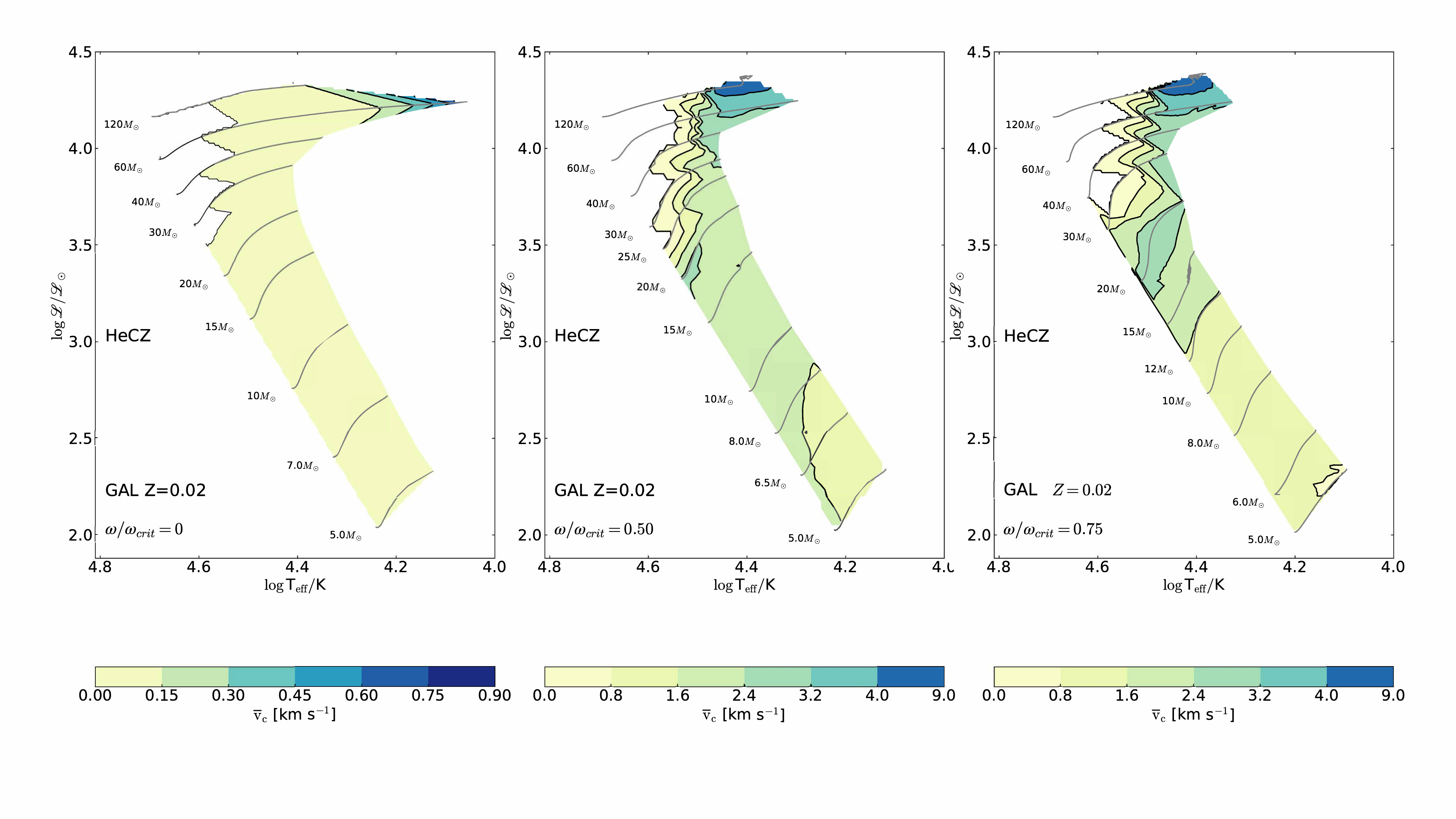}}
  \caption{
  The average convection velocities $\langle{V}_{\rm c}\rangle$ of HeCZs in the spectroscopic Hertzsprung-Russell (sHR) diagram for the models with $Z$ = 0.02 during the MS phase. The initial masses of stars are marked near evolutionary tracks. The left, middle and right panels present the models with the initial rotational velocities of $\mit\omega/\omega_{\text {crit}}$ = 0, 0.5, and 0.75, respectively. 
  The main content of the MS period is shown. The color bars show the HeCZ average convection velocity magnitude, which is calculated by Eq. ~\ref{Eq:HeCZ_aver_V_conv}. The maximum convective velocity of HeCZ in the middle panel is $\sim$ 4.8 ${\, {\rm km}\, {\rm s}^{-1}}$.}
  \label{fig:HeCZ_GAL_image}
\end{figure}

The subsurface CZs in massive stars consist of HeCZ and FeCZ. 
For the HeCZ, Fig. \ref{fig:HeCZ_GAL_image} presents the evolution of  the average convection velocities for the Galactic models with different initial masses and rotational velocity during the MS phase in the spectroscopic Hertzsprung-Russell (sHR) diagram. Here, sHR shows the inverse of the flux-mean gravity ($\mathscr{L} \equiv {\rm T}^4_{\rm eff}/g$) vs. $T\rm_{eff}$ \citep{2014Langer}.
As shown in Fig. \ref{fig:HeCZ_GAL_image}, $\langle{V}_{\rm c}\rangle$ of HeCZs increases with the enhancing the stellar rotational velocity. The maximum $\langle{V}_{\rm c}\rangle$ rises from $\sim$ $0.9 {\, {\rm km}\, {\rm s}^{-1}}$ to $\sim$ $9.0 {\, {\rm km}\, {\rm s}^{-1}}$ when the rotational velocity increases from 0 to 0.75 times of the critical velocity. 
Notably, there is an effective temperature threshold around $T\rm_{eff}$  $\gtrsim 10^{4.6}$. The left panel shows that the HeCZ begins to disappear in early MS stars with initial masses ($M_{\text {init}} \geq$ 21 \,$M_{\odot}$). The middle and right panels indicate the disappearance of the HeCZ in early MS stars at $M_{\text {init}} \geq$ 23 \,$M_{\odot}$ and $M_{\text {init}} \geq$ 30 \,$M_{\odot}$, respectively. This is attributed to the surface temperature exceeding the ionization temperature of helium, which prevents the existence of the HeCZ.

\begin{figure}
  \centering
  \resizebox{\hsize}{!}{\includegraphics{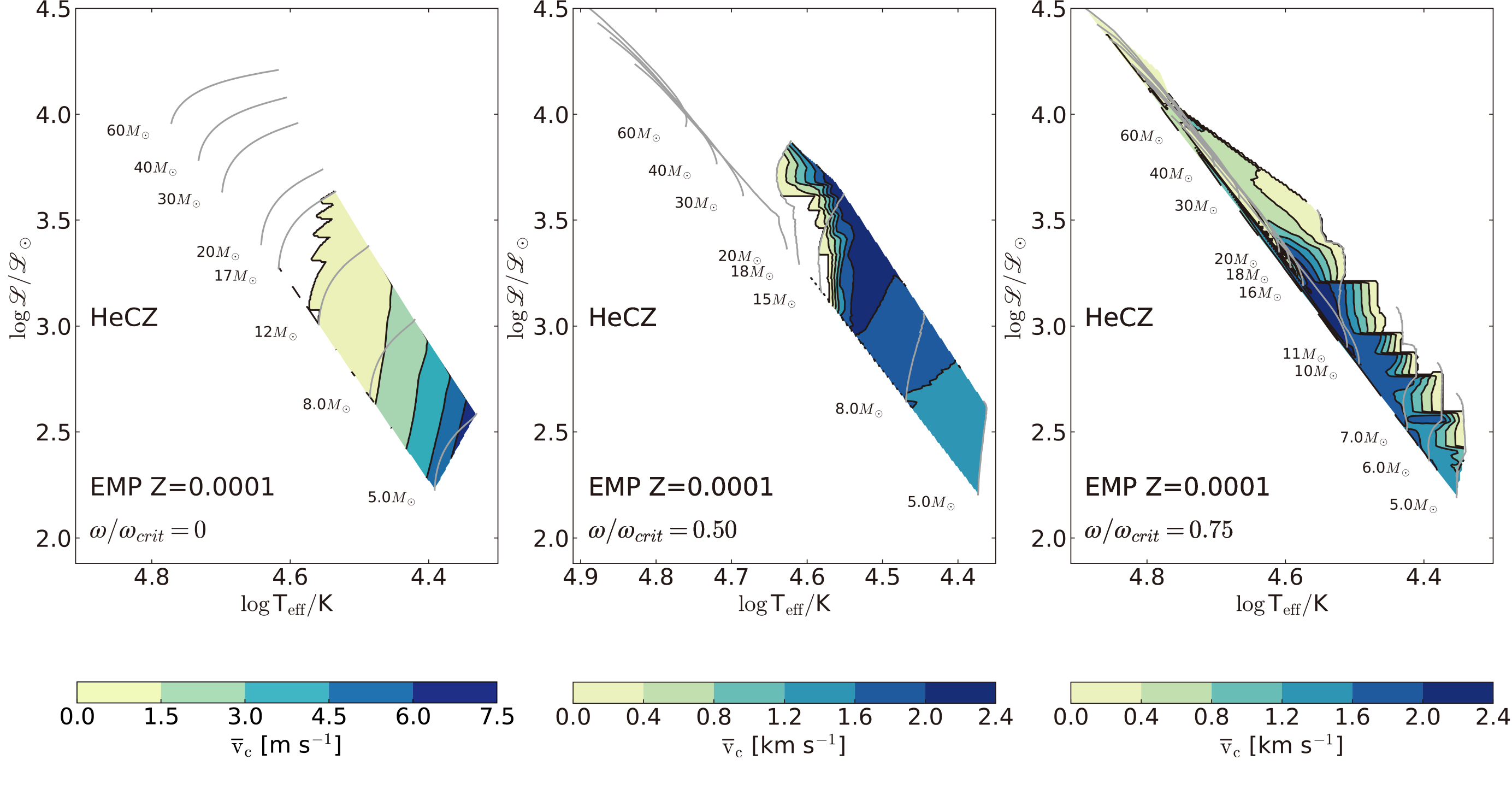}}
  \caption{
  Similar to Fig. \ref{fig:HeCZ_GAL_image}, but for the  low-metallicity models with $Z$ = 0.0001, and the initial mass between 5 \,$M_{\odot}$ and 60 \,$M_{\odot}$. The average convective velocity  is displayed in ${\, {\rm m}\, {\rm s}^{-1}}$ at the bottom of the left panel.
  }
  \label{fig:HeCZ_EMP_image}
\end{figure}

Simultaneously, similar to Fig. \ref{fig:HeCZ_GAL_image}, $\langle{V}_{\rm c}\rangle$ of HeCZ for  the stars with the extremely metal-poor ($Z=0.0001$) are given in Fig. \ref{fig:HeCZ_EMP_image}. The effects of the rotational velocity on $\langle{V}_{\rm c}\rangle$ of HeCZ become more pronounced. 
The average convective velocity has an order-of-magnitude increase between non-rotating and rotating models. 
Furthermore, rapid rotation leads to CHE in stars, shifting their evolutionary tracks toward the blue zone, resulting in the presence of HeCZ throughout the MS for all rapidly rotating stars. The main reasons is that, with increasing rotation speed, the CHE enhances the influence of the mean molecular weight gradient (${\nabla_{\!\mu}}$) in the CZ.

As shown by Figs. \ref{fig:HeCZ_GAL_image} and \ref{fig:HeCZ_EMP_image}, in the models with solar metallicity and high rotational speed, stars have high mass-loss rates, which leads to much higher angular momentum loss. The stelar rotation speed in these models rapidly decreases in shorter timescale than that of CHE. Therefore, these models do not undergo CHE \citep{2011Brott}.

\begin{figure}
  \centering
  \resizebox{\hsize}{!}{\includegraphics{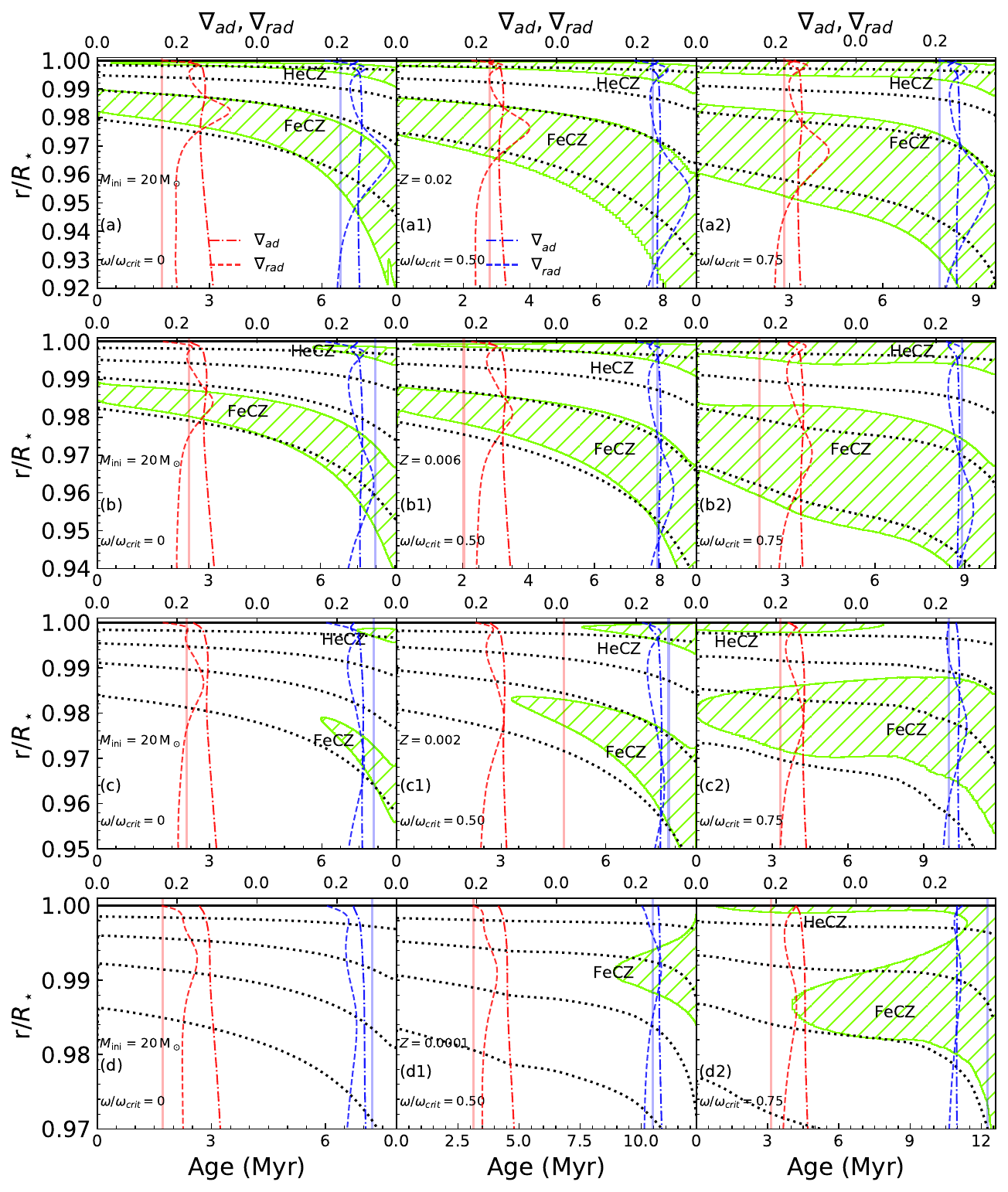}}
  \caption{
  Similar to Kippenhahn diagram of Fig. \ref{fig:20M_0.02z_0.5w_image} (the left panel), but for models with initial mass \,$M_{\text {int} }$ = 20 \,$M_{\odot}$, while having different initial rotational velocities ($\mit\omega/\omega_{\text {crit }}$ = 0, 0.5 and 0.75 in the left, middle and right panels, respectively.)  and different  metallicities ($Z$ = 0.02, 0.006, 0.002 and 0.0001 from the top to bottom panels).
  }
  \label{fig:kipp_plot_20M_image}
\end{figure}

Fig. \ref{fig:kipp_plot_20M_image} shows the effects of rotation velocities on the subsurface CZs (HeCZ and FeCZ) of 20 \,$M_{\odot}$ star with different metallicities. Obviously, the HeCZ is significantly weaker than the FeCZ, both in radial thickness and convective velocity, which is consistent with \citet{2009Cantiello}. Therefore, we focus our study on the FeCZ. With the increase of the initial rotational velocity from 0 to 0.75 $\mit\omega/\omega_{\text {crit }}$, the depth and the scope of the subsurface CZs expand. Especially, for the models with EMP, there is not any subsurface CZ for non-rotation model, but a very deep FeCZ, appears in the model with a high rotational velocity of 0.75 $\mit\omega/\omega_{\text {crit }}$. CHE is favorable to the FeCZ formation. Rapid rotation triggers CHE, which reduces chemical gradient in stellar envelope. Based on the Ledoux criterion for convection, the decrease of mean molecular weight gradiet (${\nabla_{\!\mu}}$) is conducive to the convection occurring.

As shown in Fig. \ref{fig:kipp_plot_20M_image}, the initial position of the FeCZ in the models with $Z$ = 0.02 shifts from $r/R_{\star}$ $\sim$ 0.982-0.99 (bottom of the CZ - top of the CZ) to $r/R_{\star}$ $\sim$ 0.986-0.962 with increasing rotational speed. Consistent results are found at other metallicities, showing that rotation enhances the thickness of the FeCZ (also noted in Table ~\ref{table}) and allows it to extend deeper into the star. Notably, in rapidly rotating models, the thickness of the FeCZ exceeds 5$H_{\rm P}$ to 7$H_{\rm P}$ across all metallicities. The presence of the FeCZ strictly adheres to the dynamical instability condition, the Ledoux criterion:$ \nabla_{\rm rad}$ $<$ $\nabla_{\rm ad}$ + $\frac{\varphi}{\delta}$ ${\nabla_{\!\mu}}$. Based on the dash-dotted and dashed lines, the radial range of the top and bottom of the FeCZ exceeds the $\nabla_{\rm rad} \geq  \nabla_{\rm ad}$ range, with this trend becoming more pronounced as metallicity decreases. This is due to the larger burning core in stars with rotational speed increasing \citep{2023Wang}, which enhances the radiative gradient $\nabla_{\rm rad}$ (Also see  refer to the last part of the first paragraph of Sect. 3). Simultaneously, radid rotation enhances internal mixing, making it difficult to establish a distinct chemical gradient. Thereby, high rotational speed reduces the mean molecular weight gradient ${\nabla_{\!\mu}}$. This trend becomes more evident with decreasing metallicity \citep{2011Brott}. Consequently, for the Ledoux criterion, the increase in the radiative gradient $ \nabla_{\rm rad}$ and the decrease in the mean molecular weight gradient ${\nabla_{\!\mu}}$ make it easier to satisfy the conditions for convection occurring. Here, the mean molecular weight gradient ${\nabla_{\!\mu}}$ plays a more critical role, as its reduction significantly impacts the existence of the FeCZ.

\begin{figure}
  \centering
  \resizebox{\hsize}{!}{\includegraphics{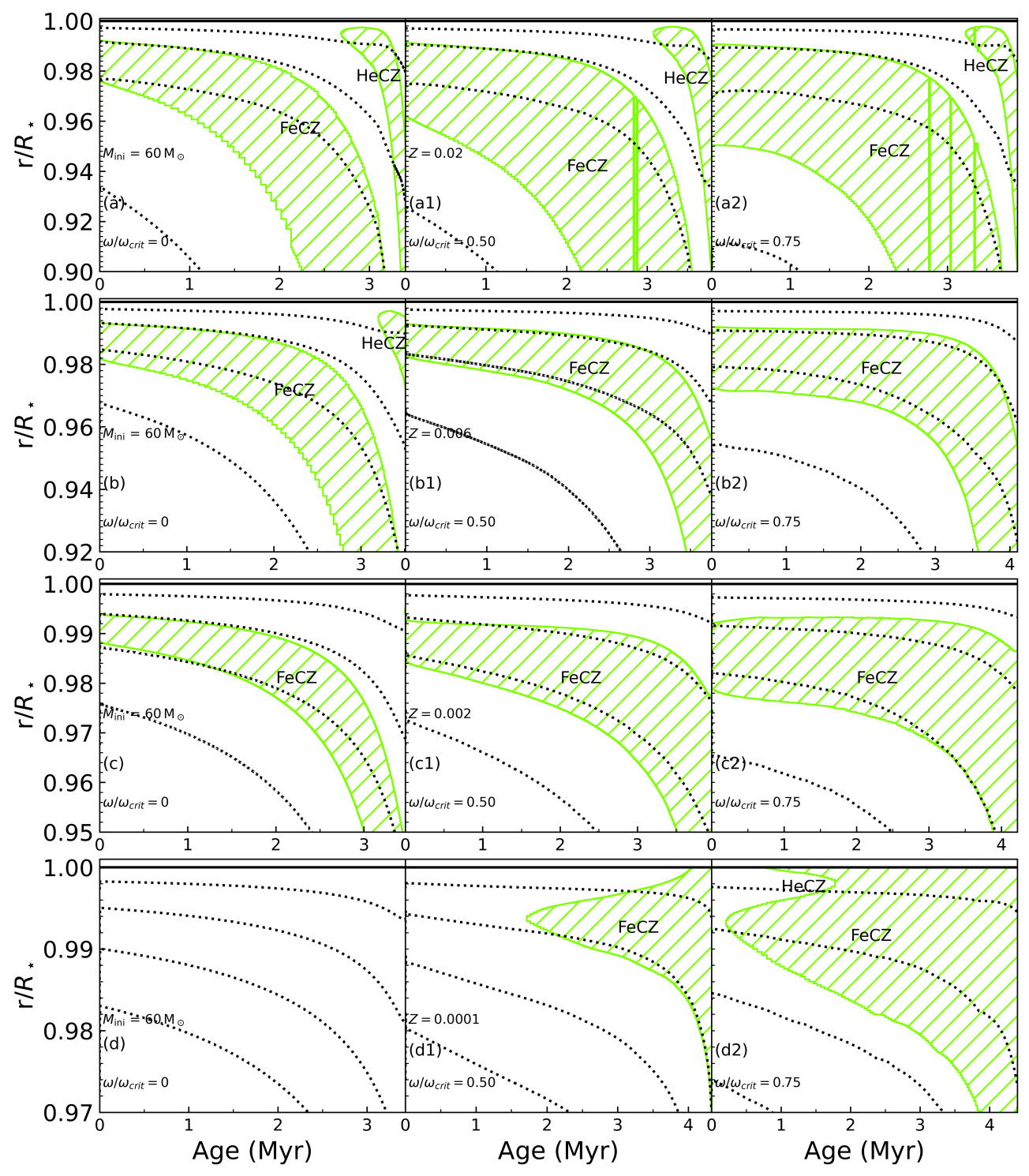}}
  \caption{
  Similar to Fig. \ref{fig:kipp_plot_20M_image}, but for a stellar model with an initial mass of $M_{\text {init} }$ = 60 \,$M_{\odot}$, and it does not include the evolution of the radiative and adiabatic gradients with radius.
  }
  \label{fig:kipp_plot_60M_image}
\end{figure}

Similarly, Fig. \ref{fig:kipp_plot_60M_image} shows the effects of rotation velocities on the subsurface CZs for more massive star with an initial mass of 60 \,$M_{\odot}$, under varying metallicities and rotation speeds. 
Compared to 20 \,$M_{\odot}$ star shown in Fig. \ref{fig:kipp_plot_20M_image}, the FeCZ of 60 \,$M_{\odot}$ star moves closer to the stellar surface, with its upper boundary near 3$H_{\rm P}$. 
The thickness of the FeCZ in more-massive star increases and extends further into the stellar interior. 
In rapidly rotating EPM models, the FeCZ and HeCZ connect together during the late MS phase, appearing at the stellar surface, where the FeCZ thickness can reach $\delta r/R_{\star}$ $\sim$ 3.3 $\%$ (In rapidly rotating EMP models with $M_{\text {init} }$ = 10 \,$M_{\odot}$, the thickness of the FeCZ can reach $\delta r/R_{\star}$ $\sim$ 7 $\%$). 
The connection of the FeCZ and HeCZ means significant surface fluctuations at this stage. In all models where the two subsurface CZs converge, CZs are confirmed to exist on the stellar surface. This interaction results in substantial radial (but not mass) fluctuations, indicating considerable pulsations during this phase \citep{2013Braithwaite,2019Cantiello,2021Balona}.

\begin{figure}
  \centering
  \resizebox{\hsize}{!}{\includegraphics{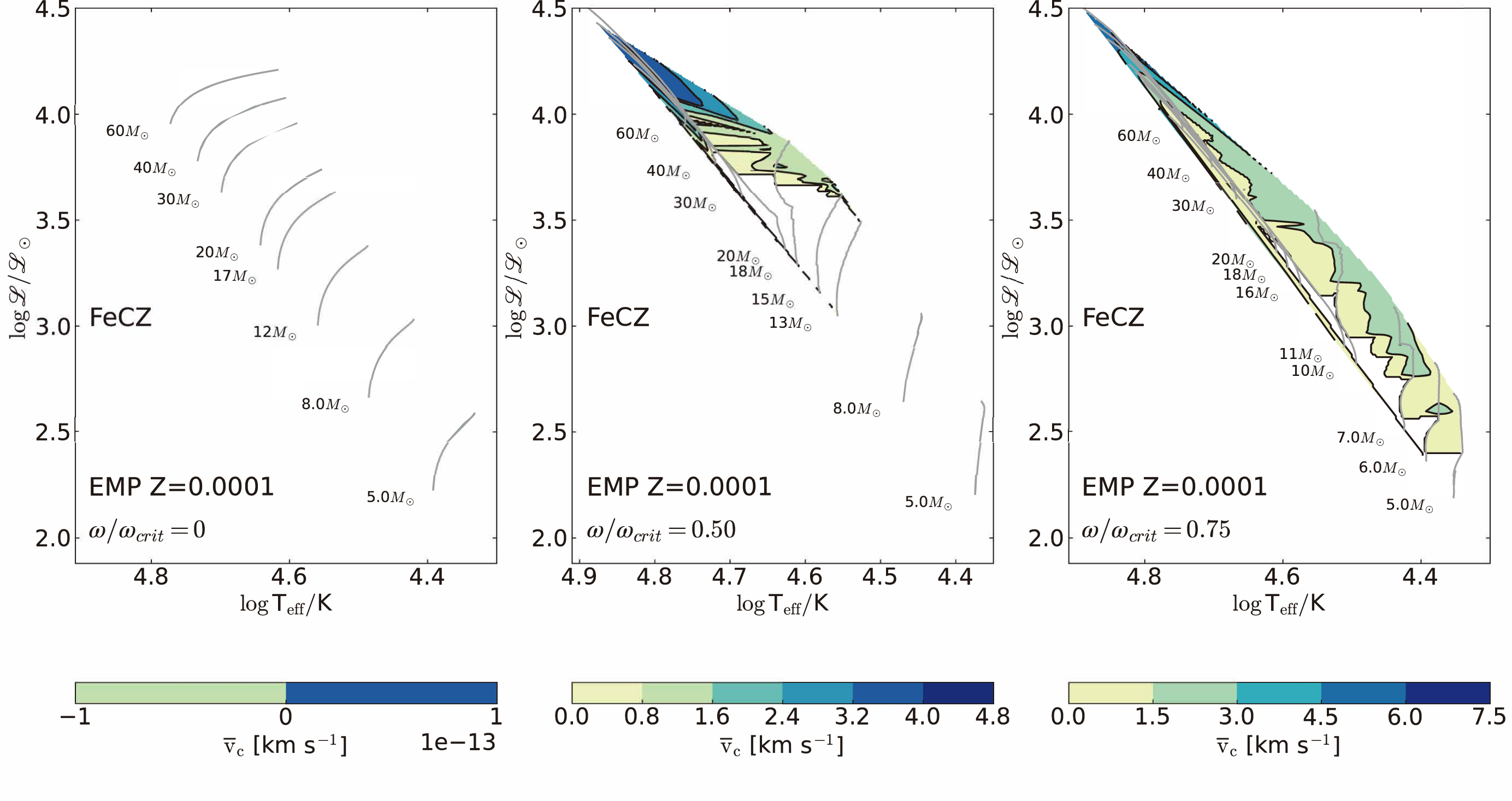}}
  \caption{
  Similar to Fig. \ref{fig:HeCZ_EMP_image}, but for the average convective velocity of the FeCZ (calculated by Eq.~\ref{Eq:FeCZ_aver_V_conv}) in EMP stars models.
  }
  \label{fig:sHR_EMP_image}
\end{figure}
To further investigate the evolution of the FeCZ in EMP stellar models during the MS phase, the average convective velocity ( $\bar{V}_{\rm c}$ ) of the FeCZ (calculated by Eq.~\ref{Eq:FeCZ_aver_V_conv}) in the sHR diagrams for three different rotation rates is shown in Fig. \ref{fig:sHR_EMP_image}. 
As Fig. \ref{fig:sHR_EMP_image} shows, there is not the subsurface CZ throughout MS in non-rotating models. This is due to the low metallicity models having lower opacity, allowing more flux to be transmitted radiatively, resulting in the absence of a subsurface CZ. In rotating models, FeCZ first appears at the end of the MS in the $M_{\text {init} }$  = 14 \,$M_{\odot}$ model when $\mit\omega/\omega_{\text {crit}}$ = 0.50, although its development duration is short within $\sim$ 0.5 Myr ($\sim$ 3.6$\%$ of the MS lifetime). 
The main reason is that the mass of central burning region increases with stellar mass and rotational speed enhance and metallicity decrease. It means that these models have high stellar luminosity, which results in high radiative gradient $\nabla_{\rm rad}$. Based on the Ledoux criterion, a large $\nabla_{\rm rad}$ triggers convection more easily. Therefore, the subsurface CZ appears in this model. 
Additionally, stellar models with initial mass $M_{\text {int}} \geq 19 \,M_{\odot}$ undergo blueward evolution due to CHE. 
It is well known that CHE makes the mixing between the nuclear burning region and stellar envelope. With the hydrogen abundance of stellar envelope decreasing, the envelope opacity also reduces. It indicates that the stellar radius becomes smaller, and its effective temperature rises. Therefore, these stars evolves toward blue in the sHR diagram.  
A spectral luminosity threshold of $\sim 10^{3.5} \mathscr{L}_\odot$ is identified, below which the FeCZ does not exist. When $\mit\omega/\omega_{\text {crit}}$ = 0.75, the FeCZ is present across the entire model range, with a maximum average convective velocity of $\sim$ 7 ${\, {\rm km}\, {\rm s}^{-1}}$, typically peaking near the end of the MS. In this set of models, the spectral luminosity threshold for the FeCZ is $\sim 10^{2.4} \mathscr{L}_\odot$.

\begin{figure}
  \centering
  \resizebox{\hsize}{!}{\includegraphics{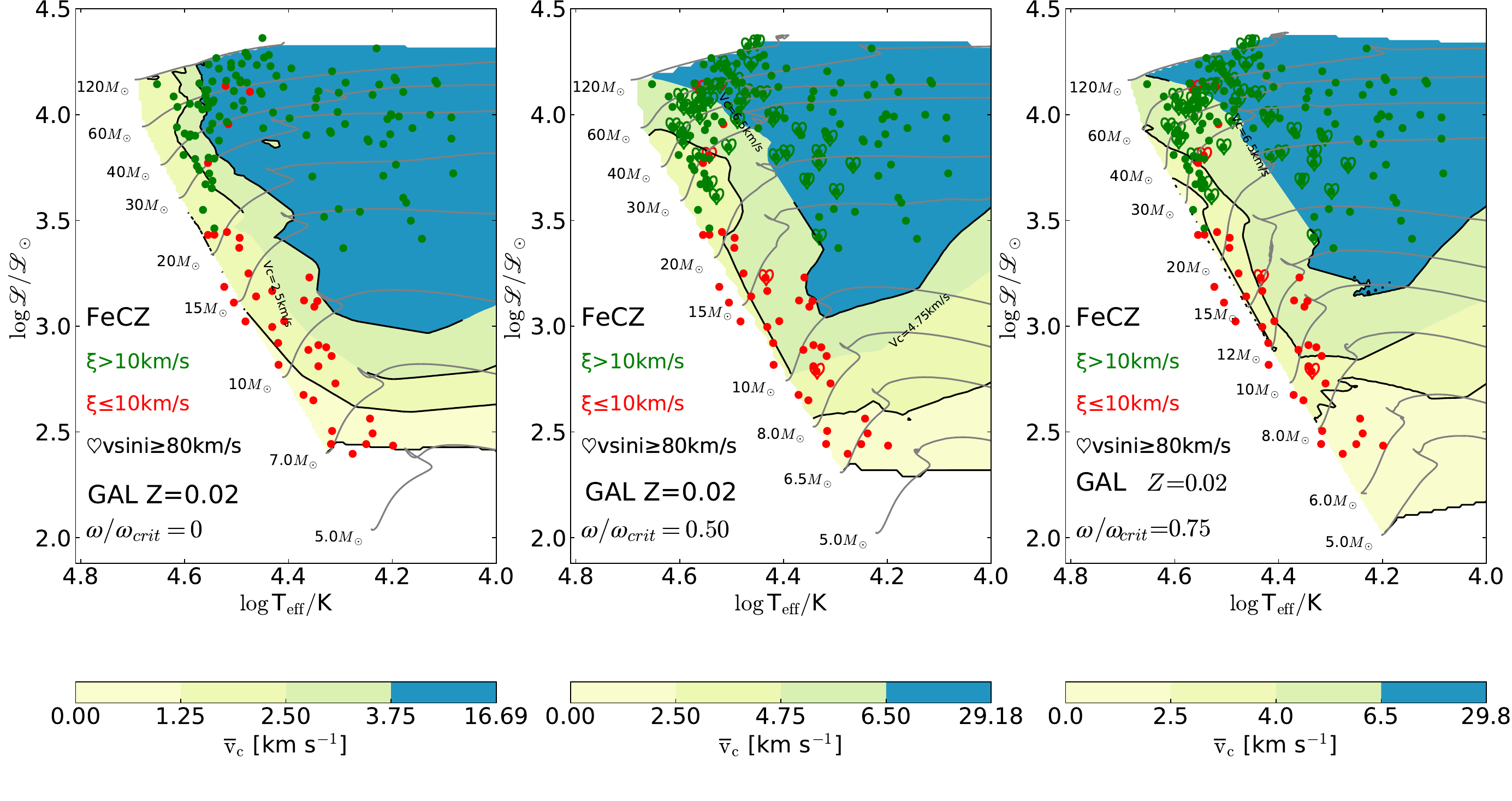}}
  \caption{
  Average convective velocity ($\bar{V}_{\rm c}$, calculated according to Eq.~\ref{Eq:FeCZ_aver_V_conv}) within 1.5 pressure scale heights ($\alpha_{\mathrm{LMT}}$$H$$_{\rm P}$) of the upper border of the FeCZ in our models, as a function of the location in the sHR diagrams (see colorbar, black contour lines), based on evolutionary tracks between 5 \,$M_{\odot}$ and 120 \,$M_{\odot}$ (gray solid lines), for metallicity corresponding to the Galaxy $Z$ = 0.02. Non-rotating models are shown in the left panel. The rotating models with $\mit\omega/\omega_{\text {crit }}$ = 0.50 and $\mit\omega/\omega_{\text {crit }}$ = 0.75 are shown in the middle and right panels. The circles represent the observed OB stars, color-coded observations based on their photospheric microturbulence velocities ${\xi}$ as provided by \citet{2007Trundle,2009Hunter,2010Fraser,2012Nieva} and \citet{2018Holgado}. Here, for the model without rotation we use only apparent rotational velocities of {$v \sin i$\,}$ < 80{\, {\rm km}\, {\rm s}^{-1}} $. We denote by heart-shaped $\heartsuit$ the observed target with an apparent rotation velocity of {$v \sin i$\,}$ \geq 80{\, {\rm km}\, {\rm s}^{-1}} $. The uncertainty in determining ${\xi}$ is typically $\pm 3\sim 5{\, {\rm km}\, {\rm s}^{-1}}$.
  }
  \label{fig:HR_GAL_image}
\end{figure}

\begin{figure}
  \centering
  \resizebox{\hsize}{!}{\includegraphics{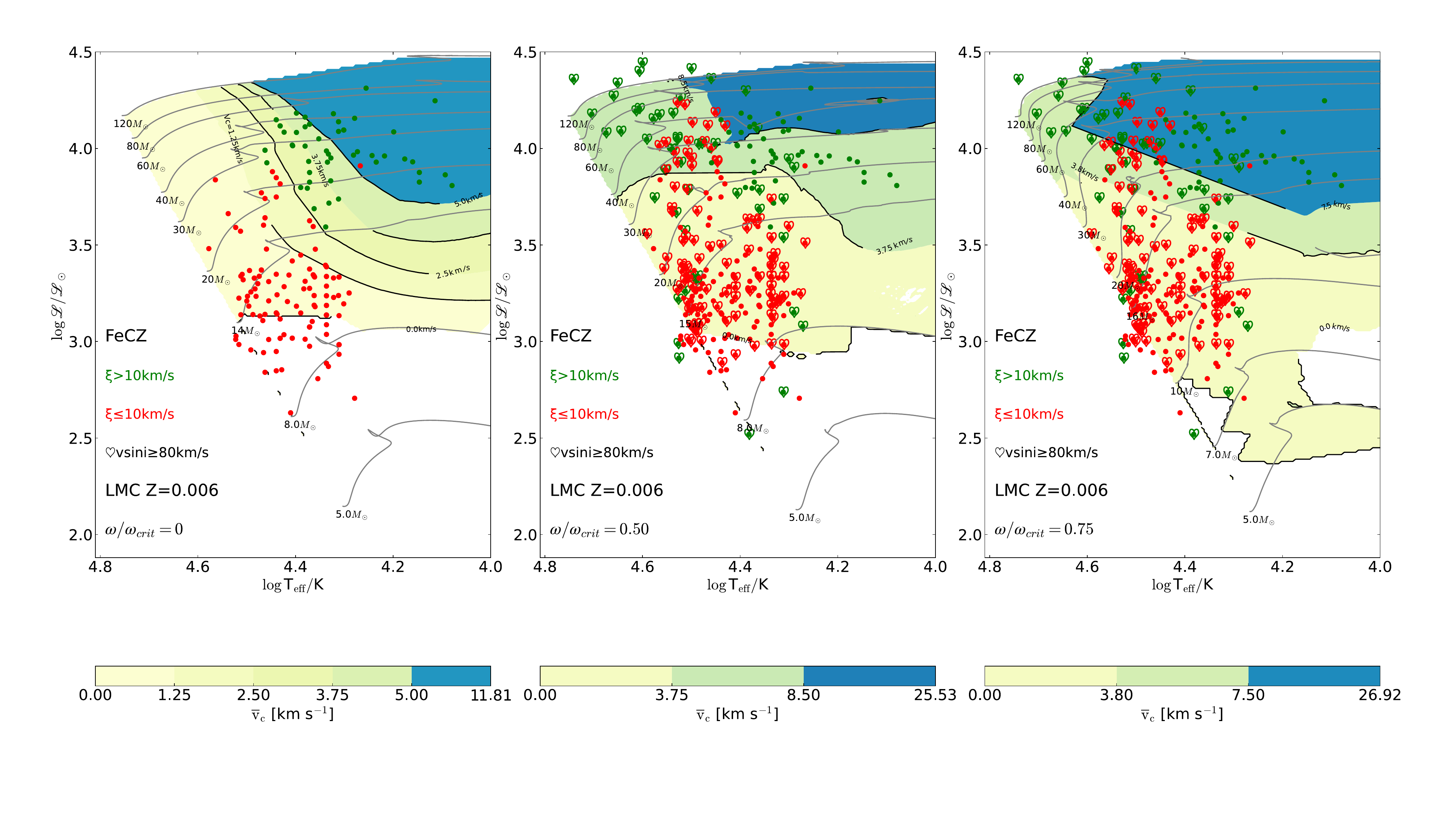}}
  \caption{
  Similar to Fig. \ref{fig:HR_GAL_image}, but for the models in the LMC with $Z$ = 0.006.
  The circles represent the observed OB stars, color-coded observations based on their photospheric microturbulence velocities ${\xi}$ as provided by
  \citet{2007Trundle,2007Hunter,2009Hunter,2011Dunstall,2015McEvoy,2017Ramrez} and \citet{2018Dufton}.
  }
  \label{fig:HR_LMC_image}
\end{figure}

In a short, the rotation has significant effects on the subsurface CZs. It can enhance the thickness of the subsurface CZs. Especially, the rapid rotation can trigger the formation of subsurface CZs in EMP stars.  

\subsection{Comparison with observations}\label{sec:three_observations}
\begin{figure}
  \centering
  \resizebox{\hsize}{!}{\includegraphics{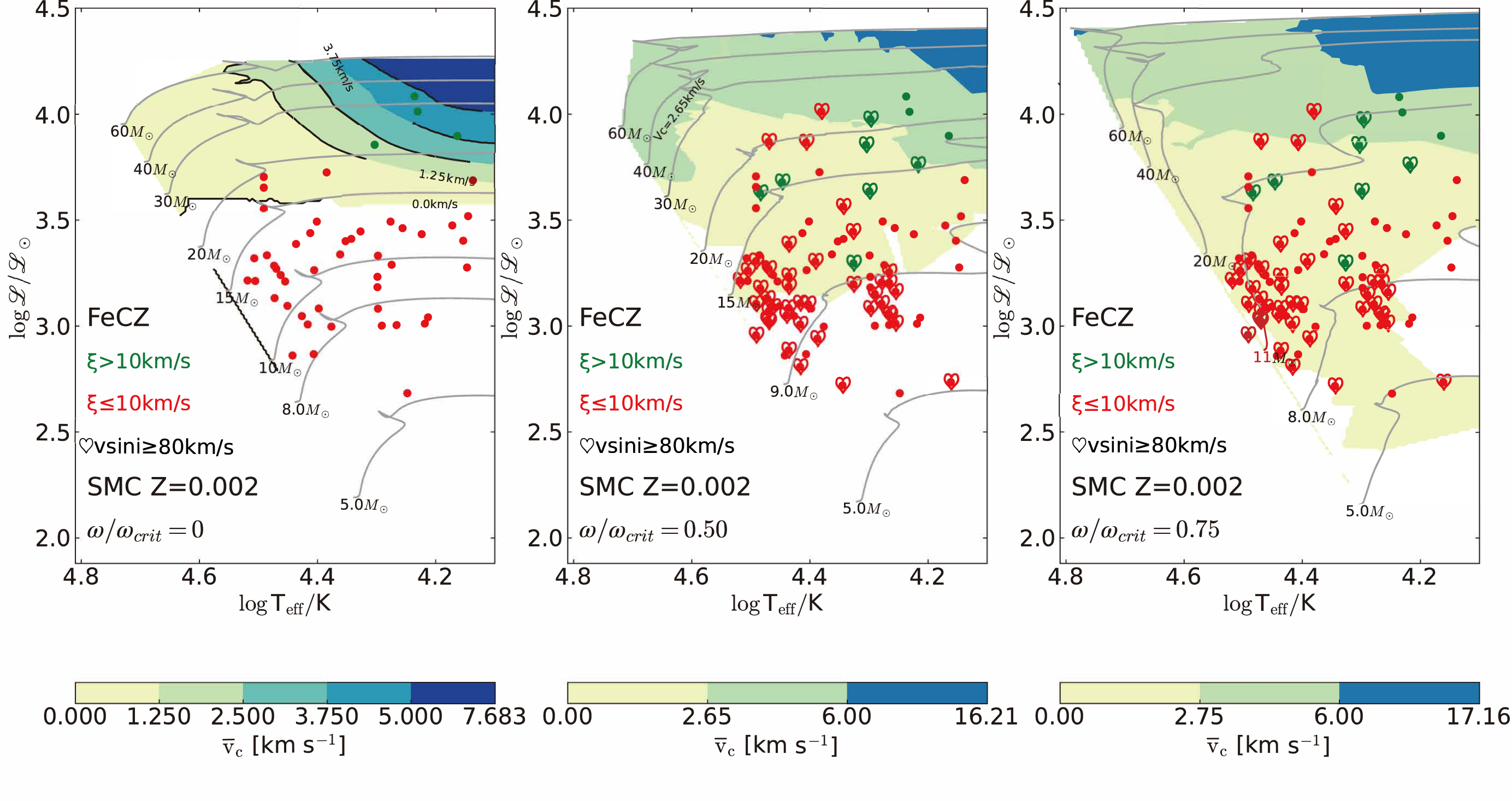}}
  \caption{
  Similar to Fig. \ref{fig:HR_GAL_image}, but for the models with $Z$ = 0.002, and the initial mass between 5 \,$M_{\odot}$ and 60 \,$M_{\odot}$. The circles represent the observed OB stars, color-coded observations based on their photospheric microturbulence velocities ${\xi}$ as provided by \citet{2007Trundle,2007Hunter,2009Hunter} and \citet{2011Dunstall}.
  }
  \label{fig:HR_SMC_image}
\end{figure}

Building on the evolution of the subsurface CZs induced by rotation, we investigate its potential role as the origin of observable phenomena, such as microturbulence at the stellar surface. We highlight results comparing microturbulence in massive stars, as presented in Figs. \ref{fig:HR_GAL_image}, \ref{fig:HR_LMC_image} and \ref{fig:HR_SMC_image} .

Convective surface velocity fluctuations can affect stellar spectra. If the surface velocity fluctuations are relevant over the entire line-forming region, it is called macroturbulence. Otherwise it is referred to as microturbulence \citep{2019Lecoanet}. Microturbulence is a form of turbulence that varies on small distance scales (scales shorter than the photon mean free range) \citep{1998Smith}, and in the chromosphere of stars, microturbulence may contribute to the broadening of absorption lines in stellar spectra \citep{2024Debnath}. Microturbulence varies with the effective temperature and surface gravity \citep{2007Hunter}. We use the classification method of \citet{2009Cantiello} to classify the observed sample into stars with significant (${\xi}$ $>$ 10 ${\, {\rm km}\, {\rm s}^{-1}}$) and insignificant (${\xi}$ $\leq$ 10 ${\, {\rm km}\, {\rm s}^{-1}}$) microturbulence.

Fig. \ref{fig:HR_GAL_image} presents the average convective velocity $\bar{V}_{\rm c}$ within a distance of $\alpha_{\mathrm{LMT}}$$H$$_{\rm P}$ from the upper boundary of the FeCZ, depicted as contour lines in the sHR diagram. 
By comparison of the three panels, we can find that in the non-rotating models, the contour for $\bar{V}_{\rm c}$ = 2.5 ${\, {\rm km}\, {\rm s}^{-1}}$ delineates a boundary between red and green points, consistent with the findings of \citet{2009Cantiello}. In the rotating models, $\bar{V}_{\rm c}$ = 6.5 ${\, {\rm km}\, {\rm s}^{-1}}$ serves as a boundary for two sets of observations. The maximum value on the color scale reflects the highest average convective velocity for this model group, indicating that the increase of rotation velocity can results in higher average velocity. Additionally, in the models with metallicity $Z$ = 0.02, we find that as rotation increases, the spectral luminosity ($\mathscr{L}$/$\mathscr{L}_\odot$) threshold of the FeCZ decreases from  $\sim 10^{2.5} \mathscr{L}_\odot$ to $\sim 10^{2.1} \mathscr{L}_\odot$. Notably, a significant portion of the observed sample with {$v \sin i$\,}$ \geq 80{\, {\rm km}\, {\rm s}^{-1}} $ clusters around $M_{\text {init}} \geq 20 \,M_{\odot}$, predominantly comprising stars with pronounced microturbulence (${\xi}$ $\leq$ 10 ${\, {\rm km}\, {\rm s}^{-1}}$).

The model results for the LMC are displayed in Fig. \ref{fig:HR_LMC_image}. Comparing the three panels, we find that in the non-rotating model, the contour for $\bar{V}_{\rm c}$ = 2.5 ${\, {\rm km}\, {\rm s}^{-1}}$ delineates a boundary between red and green points. In the panel with rotation at $\mit\omega/\omega_{\text {crit }}$ = 0.50, $\bar{V}_{\rm c}$ = 3.75 ${\, {\rm km}\, {\rm s}^{-1}}$ serves as a boundary for two observation groups. In the high-rotation model, approximately $\bar{V}_{\rm c}$ = 3.8 ${\, {\rm km}\, {\rm s}^{-1}}$ can also be considered a boundary. We note that the increased luminosity due to rotation may render the FeCZ velocity more unstable, leading to some fluctuations in this contour while maintaining an overall trend. Additionally, in the right panel, the FeCZ disappears in low-mass stars ( 5 \,$M_{\odot} \la M_{\text {init}} \la 10 \,M_{\odot}$ ) during the supergiant phase. 
The main reason is that, in the supergiant phase, the stellar radius increases and the stars cool down. The temperature of stelar subsurface fails to reach the Fe peak temperature, and the FeCZ disappears. 
The spectral luminosity of the FeCZ is noted as $\sim 10^{3.1} \mathscr{L}_\odot$, $\sim 10^{2.9} \mathscr{L}_\odot$, $\sim 10^{2.3} \mathscr{L}_\odot$, from left to right. Due to the decreasing spectral luminosity threshold, the high-rotation models encompass all observed samples. 

Fig. \ref{fig:HR_SMC_image} presents the stellar model results for metallicity $Z$ = 0.002. 
In the non-rotating models, the contour for $\bar{V}_{\rm c}$ = 1.25 ${\, {\rm km}\, {\rm s}^{-1}}$ delineates a boundary between red and green points. In the case of rotation at $\mit\omega/\omega_{\text {crit }}$ = 0.50, $\bar{V}_{\rm c}$ = 2.65 ${\, {\rm km}\, {\rm s}^{-1}}$ serves as a boundary (excluding observations with {$v \sin i$\,}$ \geq 80{\, {\rm km}\, {\rm s}^{-1}} $). 
We consider that the {$v \sin i$\,}$ \geq 80{\, {\rm km}\, {\rm s}^{-1}} $ observations result in a more dispersed distribution of stars with significant microturbulence (green points). However, due to the lowering of the spectral luminosity threshold (from $\sim 10^{3.5} \mathscr{L}_\odot$ to $\sim 10^{3.1} \mathscr{L}_\odot$), nearly half of the observational samples are covered. In the right panel, $\bar{V}_{\rm c}$ = 2.75 ${\, {\rm km}\, {\rm s}^{-1}}$ marks the boundary for red and green observations (again excluding {$v \sin i$\,}$ \geq 80{\, {\rm km}\, {\rm s}^{-1}} $). The spectral luminosity threshold for this model group is $\sim 10^{2.4} \mathscr{L}_\odot$, with all observed samples above this threshold. Notably, there is a gap in the FeCZ for low-mass stars ( 5 \,$M_{\odot} \la M_{\text {init}} \la 10 \,M_{\odot}$ ) during the supergiant phase. The reason for this phenomenon is consistent with the LMC model.

In conclusion, our non-rotating model results are consistent with those of \citet{2009Cantiello}, but we have compared our simulations with a larger set of observational samples. Under the influence of rotation, we find a reduction in the spectral luminosity threshold of the FeCZ, allowing it to encompass all observed samples, including those not covered by \citet{2009Cantiello}. Thus, we propose that the observed microturbulence on stellar surfaces originates from the subsurface CZs.

\section{Conclusions}\label{sec:four}
Using MESA code, we investigate the effects of rotation on the subsurface CZs of massive stars with different metallicities. We find that there are significant differences between subsurface CZs in rapidly rotating massive stars and those in non-rotating counterparts. HeCZ can appear on the surface of rapidly rotating massive stars, with average convective velocities ranging from 2 to 9 ${\, {\rm km}\, {\rm s}^{-1}}$. In these stars with low metallicity, the effective temperature threshold of Log $T\rm_{eff}$ $\sim$ 4.6 is no longer applicable. The rotation has significant effects on the subsurface CZs. It can enhance the thickness of the subsurface CZs. Especially, the rapid rotation can trigger the formation of subsurface CZs in EMP stars. 
Compared with HeCZ, FeCZ are dominated in rapidly rotating massive stars. 
The intensity of FeCZ increases with higher metallicity, luminosity, and rotational velocity. The emergence and positioning of FeCZ in these stars must consider the average molecular weight gradient ${\nabla_{\!\mu}}$, as rotation induces chemical homogeneity. The average convective velocity of FeCZ in rapidly rotating massive stars ranges from approximately 7 to 30 ${\, {\rm km}\, {\rm s}^{-1}}$. Moreover, rapid rotation effectively lowers the luminosity threshold for FeCZ, allowing them to manifest in stars with smaller initial masses.

Rapid rotation can reduce the threshold of the spectral luminosity of the FeCZ, which results in large microturbulence. This is consistent with the observations. It confirms the potential existence of an FeCZs origin for the observed microturbulence on the surfaces of OB stars. 
It means a potential link to the subsurface CZ origins of stochastic low-frequency variability observed in massive stars \citep{2019Lecoanet,2024Shen}.

\section*{Acknowledgements}
We thank Professor Cheng Cheng for her assistance. This work received the generous support of 
the National Natural Science Foundation of 
China under grants No.U2031204, 12163005, 12373038, 12288102 and 12263006, the 
science research grants from the China Manned Space Project with No. CMSCSST-2021-A10, and 
the Natural Science Foundation of Xinjiang No. 2022D01D85 and 2022TSYCLJ0006,  
and the Major Science and Technology Program of Xinjiang Uygur Autonomous Region under grant No.2022A03013-3.

\bibliographystyle{raa}
\bibliography{bibtex}

\end{document}